\documentclass[12pt,letterpaper]{article}
\usepackage{osajnl}
\usepackage{endfloat}
\usepackage{xspace}
\newcommand{\citep}[1]{\cite{#1}}
\newcommand{\ignore}[1]{}
\newcommand{\Ao}{Adaptive optics (AO)\renewcommand{\Ao}{AO\xspace}\renewcommand{\ao}{AO\xspace}\xspace}
\renewcommand{\ao}{adaptive optics (AO)\renewcommand{\ao}{AO\xspace}\renewcommand{\Ao}{AO\xspace}\xspace}
\newcommand{\elts}{extremely large telescopes (ELTs)\renewcommand{\elts}{ELTs\xspace}\renewcommand{\elt}{ELT\xspace}\xspace}
\newcommand{\elt}{extremely large telescope (ELT)\renewcommand{\elts}{ELTs\xspace}\renewcommand{\elt}{ELT\xspace}\xspace}
\newcommand{\mcao}{multi-conjugate AO (MCAO)\renewcommand{\mcao}{MCAO\xspace}\xspace}
\newcommand{\xao}{extreme AO (XAO)\renewcommand{\xao}{XAO\xspace}\xspace}

\newcommand{\fpga}{field programmable gate array (FPGA)\renewcommand{\fpga}{FPGA\xspace}\renewcommand{\fpgas}{FPGAs\xspace}\xspace}
\newcommand{\fpgas}{field programmable gate arrays (FPGAs)\renewcommand{\fpga}{FPGA\xspace}\renewcommand{\fpgas}{FPGAs\xspace}\xspace}
\newcommand{\fwhm}{full width half maximum (FWHM)\renewcommand{\fwhm}{FWHM\xspace}\xspace}
\newcommand{\hdl}{hardware description language (HDL)\renewcommand{\hdl}{HDL\xspace}\xspace}
\newcommand{\ffts}{fast Fourier transforms (FFTs)\renewcommand{\ffts}{FFTs\xspace}\xspace}
\newcommand{\shwfs}{Shack-Hartmann wavefront sensor (SHWFS)\renewcommand{\shwfs}{SHWFS\xspace}\xspace}
\newcommand{\psfs}{point spread functions (PSFs)\renewcommand{\psfs}{PSFs\xspace}\renewcommand{\psf}{PSF\xspace}\xspace}
\newcommand{\psf}{point spread function (PSF)\renewcommand{\psfs}{PSFs\xspace}\renewcommand{\psf}{PSF\xspace}\xspace}
\newcommand{\mpi}{message passing interface (MPI)\renewcommand{\mpi}{MPI\xspace}\xspace}
\newcommand{\Glao}{Ground layer \ao (GLAO)\renewcommand{\Glao}{GLAO\xspace}\renewcommand{\glao}{GLAO\xspace}\xspace}
\newcommand{\glao}{ground layer \ao (GLAO)\renewcommand{\Glao}{GLAO\xspace}\renewcommand{\glao}{GLAO\xspace}\xspace}
\newcommand{\wfs}{wavefront sensor (WFS)\renewcommand{\wfs}{WFS\xspace}\xspace}

\newcommand{\mnras}{MNRAS}
\newcommand{\aap}{A\&A}
\newcommand{\pasp}{Pub. Astron. Soc. Pacific}

\begin{document}
\title{The Durham ELT adaptive optics simulation platform}
  \author{Alastair Basden, Timothy Butterley, Richard Myers, and
  Richard Wilson}
\address{Centre for Advanced Instrumentation, Department of Physics, Durham
  University, South Road, Durham, DH1 3LE}
\email{a.g.basden@durham.ac.uk}
\date{2006}

\maketitle
\begin{abstract}
Adaptive optics systems are essential on all large telescopes where
image quality is important.  These are complex systems with many
design parameters requiring optimisation before good performance can
be achieved.  The simulation of adaptive optics systems is therefore
necessary to categorise the expected performance.  This paper
describes an adaptive optics simulation platform, developed at Durham
University, which can be used to simulate adaptive optics systems on
the largest proposed future extremely large telescopes (ELTs) as well
as current systems.  This platform is modular, object oriented and has
the benefit of hardware application acceleration which can be used to
improve the simulation performance, essential for ensuring that the
run time of a given simulation is acceptable.  The simulation platform
described here can be highly parallelised using parallelisation
techniques suited for adaptive optics simulation, whilst still
offering the user complete control while the simulation is running.
Results from the simulation of a ground layer adaptive optics system
are provided as an example to demonstrate the flexibility of this
simulation platform.
\end{abstract}
\section{Introduction}
\Ao is a technology widely used in optical and infra-red astronomy,
and almost all large science telescopes have an \ao system.  A large
number of results have been obtained using \ao systems which would
otherwise be impossible for seeing-limited
observations\cite{2004A&A...417L..21G,2005ApJ...625.1004M}.  New \ao
techniques are being studied for novel applications such as wide-field
high resolution imaging \citep{2004SPIE.5490..236M} and extra-solar
planet finding \citep{2004ASPC..321...39M}.

The simulation of an \ao system is important as it helps to determine
how well the \ao system will perform.  Such simulations are often
necessary to determine whether a given \ao system will meet its design
requirements, thus allowing scientific goals to be met.  Additionally,
new concepts can be modelled, and the simulated performance of
different \ao techniques compared\cite{2005MNRAS.357L..26V}, allowing
informed decisions to be made when designing or upgrading an \ao
system and when optimising the system design parameters.

A full end-to-end \ao simulation will typically involve several stages
\citep{2005MNRAS.356.1263C}.  Firstly, a representation of the
atmospheric turbulence is produced, typically by generating simulated
atmospheric phase screens, often using several different screens
representing turbulence at different atmospheric heights.  The
aberrated complex wave amplitude at the telescope aperture is then
generated by modelling this atmospheric phase as seen from the
telescope pupil.  For a stratified atmospheric model, this will
involve propagating the atmospheric phase screens across the pupil, to
simulate the effect of the relative velocity of different atmospheric
layers.  The wavefront at the pupil is then passed to the simulated
\ao system, which will typically include one or more wavefront sensors
and deformable mirrors and a feedback algorithm for closed loop
operation.  Additionally, one or more science \psfs as corrected by
the \ao system are calculated.  Information about the \ao system
performance is computed from the \psfs, including quantities such as
the Strehl ratio and encircled energy.

The computational requirements for \ao simulation scale rapidly with
telescope size, and so simulation of the largest telescopes cannot be
done without special techniques, some of which follow:

\begin{enumerate}
\item Multiprocessor parallelisation
\citep{2004SPIE.5490..705L,2003SPIE.5169..218A} allows computations to
be spread across multiple processors, though can suffer from data
bandwidth bottlenecks, as often data cannot be transferred between
processors at a rate sufficient to keep them processing for a large
proportion of the time.
\item The use of dedicated hardware for algorithm acceleration
\citep{basden4} can produce large performance improvements, though is
somewhat inflexible.
\item Analytical models can also be used \citep{2003SPIE.4840..393C},
and these can give rapid results, though are not able to represent
noise sources easily.
\end{enumerate}

We here describe the approaches that we have taken to implement an
efficient and scalable simulation framework.

\subsection{The Durham adaptive optics simulation platform}
At Durham University, we have been developing \ao simulation codes for
over ten years \citep{1995SPIE.2534..265D}.  The code has recently
been rewritten to take advantage of new hardware, new software
techniques, and to allow much greater scalability for advanced
simulation of \ao systems for \elts, including \mcao and \xao systems
\citep{2004SPIE.5382..684R}.

The Durham \ao simulation platform uses the high level programming
language Python (currently, Python 2.4), to select and link together C
(ANSI standard with GCC version 3.3), Python and hardware accelerated
algorithms, as well as third party modules, giving a great deal of
flexibility.  This allows us to rapidly prototype and develop new and
existing \ao algorithms, and to prepare new \ao system simulations
quickly using Python code.  The use of C and hardware algorithms
ensures that processor intensive parts of the simulation platform can
be implemented efficiently.  The C and Python algorithms make use of
optimised libraries including FFTW (versions 2 and 3), the AMD core
math library (version 3.5, for use on AMD platforms, including BLAS
and LAPACK routines), the GNU scientific library (currently version
0.7), and the MPICH library (optimised for the Cray XD1).  This ensures
that high performance can be achieved for computationally intensive
algorithms.  The hardware accelerated algorithms are implemented
within \fpgas, which can be programmed to provide impressive
performance improvements over a standard software implementation.  The
VHDL hardware description language is used to program the \fpgas,
using the Xilinx ISE 7.1 compiler.

The simulation software will run on most Unix-like operating systems,
including Linux and Mac~OS~X.  The simulation platform hardware at
Durham consists of a Cray XD1 supercomputer \citep{crayXD1}, which
contains reprogrammable hardware for application acceleration as well
as six dual Opteron processor nodes each with 8~GB ram.  Additionally,
a distributed cluster of conventional Unix workstations is connected
by giga-bit Ethernet.  For most simulation tasks, only the XD1 is
required, though for large models, or when multiple simulations are
run simultaneously, the entire distributed cluster can be used.  The
simulation is programmed intelligently to make use of optimised
libraries and hardware acceleration when these are available, and to
use default library replacements when not (for example, the AMD core
math library is not available on a Mac~OS~X platform).

The simulation is object orientated, with high level objects (for
example a phase screen generation object, and a wavefront sensing
object) being connected together, allowing data to be passed between
them in a direction described by the user (for example, atmospheric
phase screens may be passed to a deformable mirror object).  The high level
simulation objects can contain instances of lower level objects, which
are internal to the simulation objects, and used during calculations,
for example a telescope pupil mask object, used to define which parts
of the atmospheric phase screens are sampled by the wavefront sensor.

\section{ELT simulation requirements}
When attempting to create a realistic simulation of an \ao system on
an \elt, a large amount of computing power, memory and bandwidth will
be required.  The Durham simulation platform provides these
requirements by implementing several key technologies.

\subsection{Multiple processor simulation platform}
The Durham simulation platform allows a simulation to be comprised of
multiple processes, meaning that different parts of the simulation can
run on different processors, and even different computers.  This
however means that communication between the processes is essential.
To maximise the efficiency of the simulation, we use a combination of
shared memory access (where processes have access to the same memory,
e.g.\ within a symmetric multiprocessor (SMP) system), and \mpi
communications where appropriate, and a simulation user has control
over the type of communication used.

\subsubsection{Shared memory access}
\begin{figure}[htbp]
\centerline{\includegraphics[width=6cm]{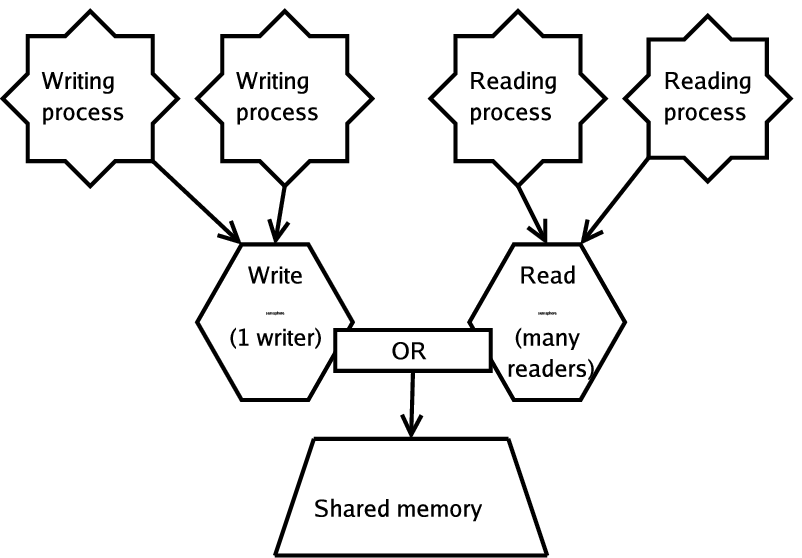}}
\caption{A schematic diagram showing how a typical shared memory
  system will operate.  Some processes will have read and write access
  to the memory, while others will have read access only.
  Synchronisation primitives will be used to ensure that data is not
  read while it is being written and vice versa.}
\label{fig:shm}
\end{figure}

Shared memory access allows multiple processes to access the same
region of computer memory.  All processes can usually have read and
write access to this memory.  Using shared memory allows a single
memory block to be shared between processes, thus reducing the overall
memory requirements, and also reducing the processor overhead, as
producing an identical copy of the data for each different processes
is then not essential.  Fig.~\ref{fig:shm} is a schematic diagram
showing how a typical shared memory system can operate.  Shared memory
buffers are created using the Unix shm\_open() function call, and are
mapped into a processors virtual address space.  Standard
synchronisation primitives, such as semaphores are used to ensure that
no processes are reading the shared memory region whilst it is being
written to, and to ensure that only one process can write to the
shared memory region at once.

The Durham simulation platform hides the use of synchronisation
primitives (in this case, semaphores) from the user (and simulation
objects), such that the parallel processes will read and write to the
shared memory region only when it is appropriate to do so.  This
removes the possibility of data corruption, whilst providing a
simplified interface for the simulation programmer.

\subsubsection{MPI communication}
Communication between distributed systems which do not share memory
requires copies of datasets to be passed between the systems.  When
the dataset is large, or when a large number of small datasets are
passed, a bottleneck can occur as processes will then spend a
significant amount of time waiting for a dataset to arrive or be sent.
It is therefore essential that the communication method used to
transfer these datasets is as efficient as possible, having both a low
latency (so that time is not wasted when sending small datasets), and
a high sustained bandwidth (so that large datasets can be sent in a
minimum time).

The Durham simulation platform uses the \mpi library for this
communication, as this allows data to be passed efficiently with only
a small latency, particularly on the XD1 system.  The Cray XD1 has an
optimised version of the \mpi library which is targeted to the
hardware architecture of this system, making efficient use of the
RapidArray Transport interface, the commercial high bandwidth
interconnect found in XD1 systems.  Using the Durham system, we have
measured the \mpi communication latency of only 1.6~$\mu$s, and a
maximum sustained bandwidth of 1.4~GBs$^{-1}$ between the computing
nodes.
  
\subsubsection{Process parallelisation}
Each processor used for a given simulation will be given only one
process to run, to reduce context switching delays.  Each of these
processes will contain one or more simulation objects, which are able
to access the virtual memory space of other objects within the
process, making data transfer between these objects trivial (e.g.\ the
address of a data array can be passed).  All simulation objects are
executed in a single thread, carrying out their computations for each
iteration in turn, which again reduces context switching delays.

Objects in separate processes are able to pass data using either \mpi
communications or shared memory as appropriate.  When such
communication is required, a pair of high level communication objects
are created and are responsible for dealing with a particular
communication link (\mpi or shared memory).  These communication
objects are then connected to the simulation objects, which can then
behave as if they are connected directly to the object with which they
wish to communicate.  Each simulation object has a basic set of
methods and data objects which are viewable by other objects.  The
communication objects then merely have to implement these methods and
data objects, transferring data as appropriate.  The use of
communication objects is transparent to the simulation objects, being
handled by the simulation framework.

\subsection{Hardware acceleration}
The Durham \ao simulation platform is able to accelerate specific
parts of the \ao simulation by using reconfigurable logic hardware,
\fpgas, and hence reducing the time taken for a given simulation to
complete.  These \fpgas are an integrated part of the Cray XD1
supercomputer \citep{basden4} and when used correctly, are capable of
reducing the execution time of some algorithms by two to three {\bf
orders of magnitude}\citep{basden6}, whilst at the same time, freeing
the CPU for other operations.  This greatly improves the speed at
which the simulation can run, and is essential for simulation of large
\ao systems.  Implementing algorithms within the \fpgas requires
knowledge of a hardware programming language, and so we have developed
common libraries which can be plugged in to an existing simulation,
for example a wavefront sensor pipeline.  The simulation user
therefore requires no hardware knowledge, and yet can achieve
significant impressive performance improvements using the hardware
acceleration.

\subsection{Simulation creation}
A user creates a new simulation by selecting and linking together the
various simulation modules as required, either graphically or in a
text file.  New modules (for example to investigate a new type of
wavefront sensor or deformable mirror) can easily be created and added
to the simulation with minimal effort.  Once the simulation file has
been set up, a parameter file is created which contains all variables
and configuration objects required by the simulation.  This parameter
file is in XML format and allows embedded Python code which can be
used to create complicated variables and objects.  If suitably
defined, a cross-simulation parameter file could be created using a
Python parser for the XML.  The parameter file can be created using a
graphical interface, which has the capability to automatically create
a skeleton parameter file from the simulation file, and then allow the
user to adjust the default values of variables.  This allows a new
simulation to be set up quickly by an inexperienced user.

\subsection{Simulation control}
Control of a running simulation is achieved by connecting to it using
either the Python command-line or graphical tools.  This gives the
user complete control over a simulation, allowing them to stop, start
and pause, as well as analyse (allowing the user to create plots of
parts of the data chain, for example, sub-aperture images) and change
the current state of a simulation (for example, changing the value of
a variable or the contents of an array).  This high degree of
flexibility is achieved by allowing the user to send text strings to
the simulation, which are treated as Python code, and executed as a
separate thread which has access to the global name-space.  The user
can therefore access and alter any part of the simulation, and any
requested data can be returned to the user for further analysis.  When
a simulation is comprised of more than one process, the user can
connect to any or all of these processes.

\begin{figure}[htbp]
\centerline{\includegraphics[width=6cm]{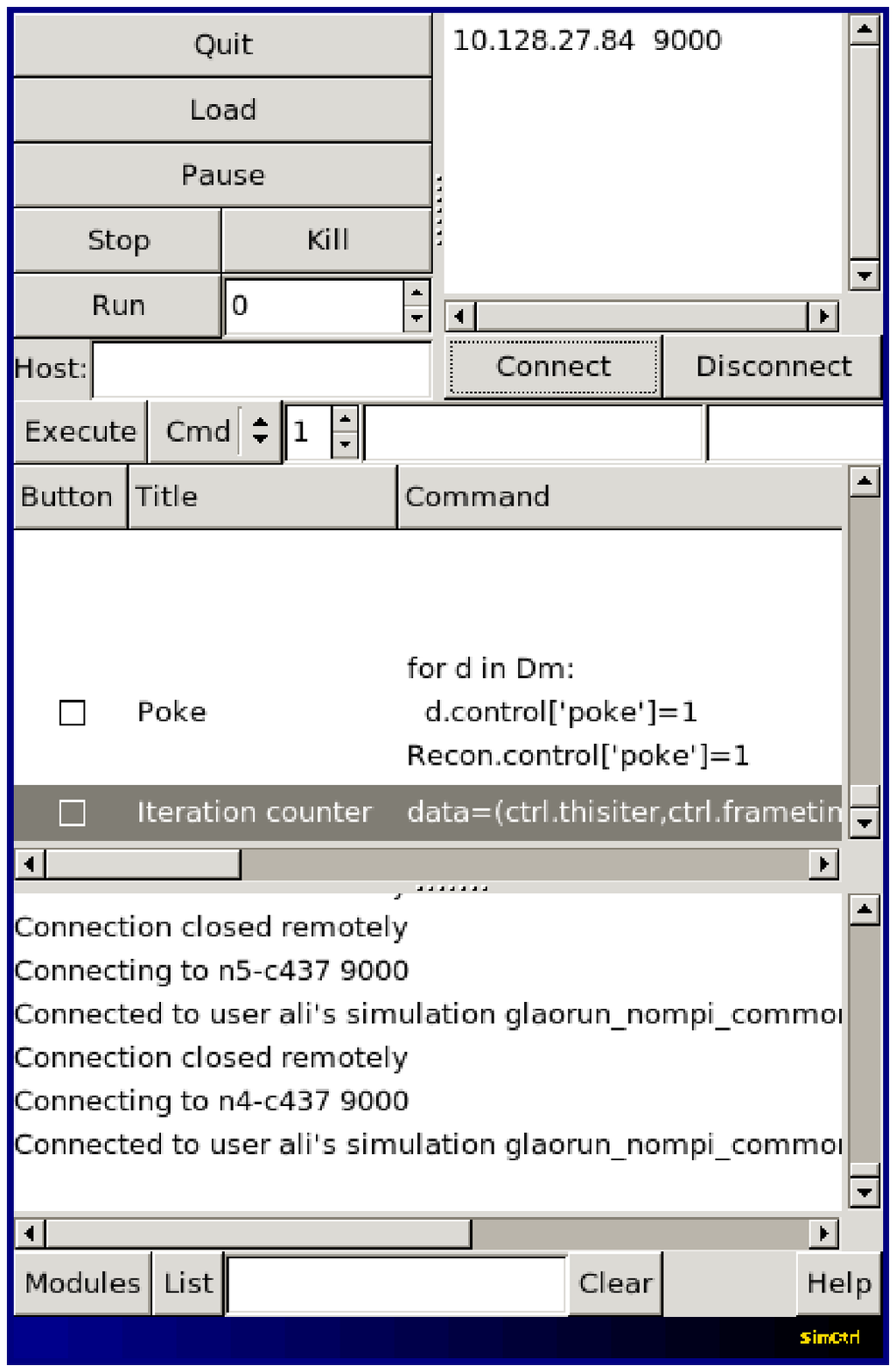}}
\caption{A screen-shot of the simulation control user interface.
  Novice users are able to control a simulation at the click of a
  button, while experienced users are able to query the simulation,
  obtain and display data, and alter the simulation state, including
  changing values and array contents.}
\label{fig:simctrl}
\end{figure}

This control facility is completely detachable from the simulation,
and can be started and stopped without affecting simulation operation.
It is also possible to have several users connected to the same
simulation at any given time, from anywhere that has Internet access
to the computers running the simulation.  Fig.~\ref{fig:simctrl} shows
a screen-shot of the simulation control user interface, and
demonstrates the powerful functionality that this provides through a
simple interface, satisfying both novice and experienced users.

This simulation control capability is unique as it enables a user to
implement new capability within a running simulation, and to query all
objects and variables, even if it was not envisaged that these should
be queried before the simulation was created.  This high degree of
flexibility is essential for \elt \ao system simulation as simulation
run-times can typically be measured in days.

\subsection{Parallelisation approaches}
\label{sect:parallel}
When parallelising any software, there is usually a trade off between
the amount of processing done, and the amount of data that has to be
passed between processors.  A bottleneck may occur if the CPUs spend a
significant amount of time waiting for data, meaning that the
parallelisation has not been efficient.  

\begin{figure}[htbp]
\centerline{\includegraphics[width=8cm]{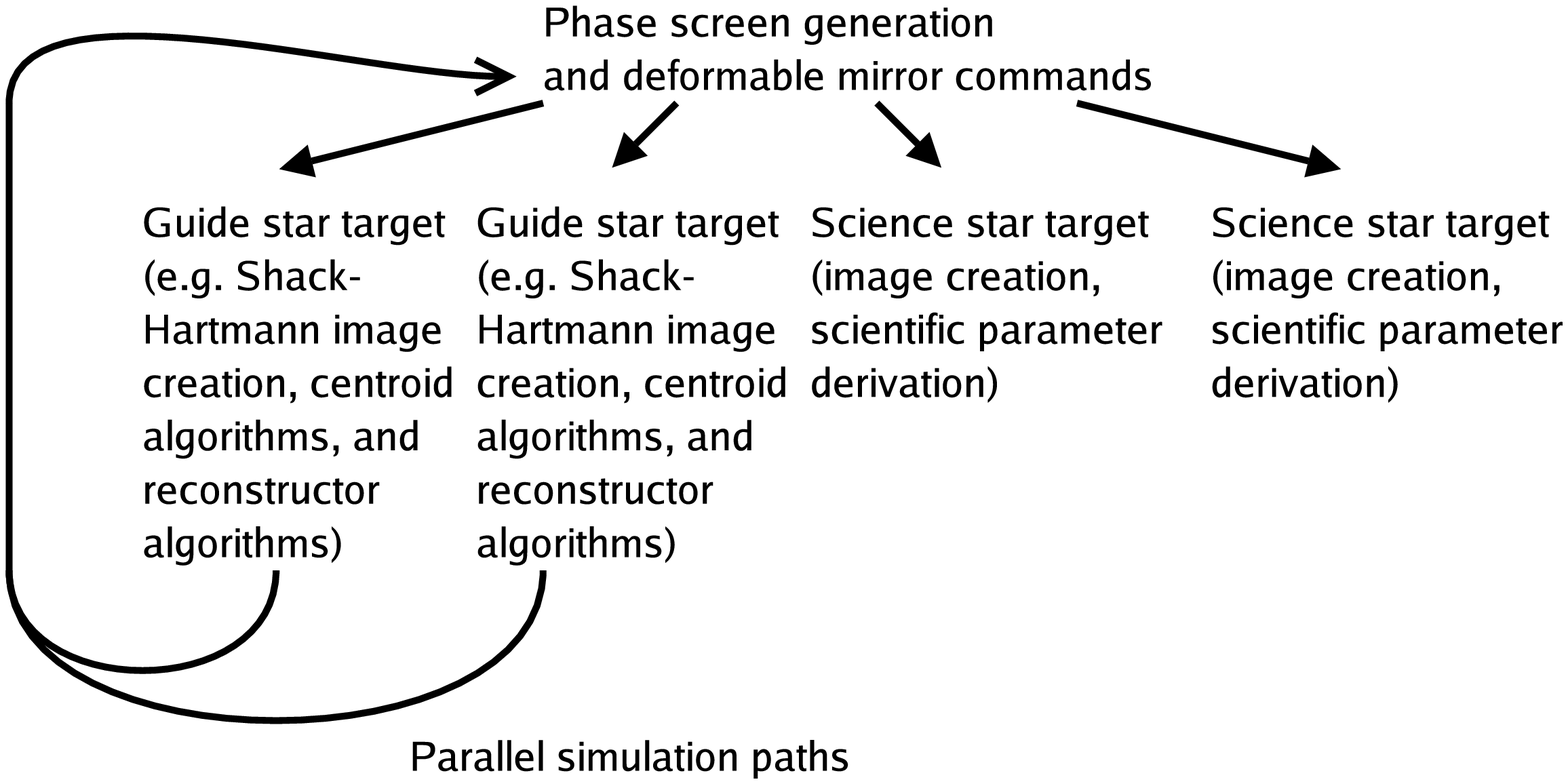}}
\caption{An example of the parallelisation of parallel optical paths.
No data-flow is required between these paths, except for the initial 
phase screens, meaning that minimal time is spent with the processors
waiting for data to arrive.}
\label{fig:opticalpath}  
\end{figure}

It is usually most efficient to create parallelised software which
sends as little data as possible between processes so that most time
can be spent processing data.  At Durham, we typically parallelise our
\ao system simulations by dividing parallel optical paths into
separate processes, as shown in Fig.~\ref{fig:opticalpath}.  Each
optical path is virtually independent of the others, except that they
all require inputs of atmospheric phase screen data and knowledge of
any time varying deformable mirror surface shapes, and may (if part of
the wavefront correction path) return new deformable mirror commands,
or wavefront sensed values to be passed to other optical paths.  By
dividing the processes in this way, a minimum amount of time is spent
waiting for data, allowing the most efficient use of the CPUs to be
made.  This will also allow a typical simulation (with several on and
off-axis science targets, and one or more guide stars) to be
parallelised into a similar number of processes as there are
processors, allowing a single process to run on each processor.

When all parallel optical paths depend on one algorithm which
generates data for the paths, for example, atmospheric turbulence
generation, or reconstruction of the deformable mirror commands from
the wavefront sensor data, this algorithm can also be parallelised
using a traditional parallelisation approach, by splitting the
computation over available processors, and passing the data as
required.  Some optimised libraries, for example the FFTW Fourier
transform library, use this technique.  However, this parallelisation
approach is only beneficial for algorithms where the time spent
transferring data is small compared with the time spent processing the
data.

\subsubsection{Simulation scalability}
To demonstrate the scalability of the \ao simulation, we have
simulated a system with three wavefront sensors ($32\times32$
sub-apertures each), one science target, and assume that atmospheric
turbulence is concentrated in two layers.
Table~\ref{tab:scaleObjects} provides details of the different
simulation objects required for this simulation, and gives typical
computation times for this example.  It should be noted that the
computation times do not scale identically with simulation size, and
so the ratio of computation times between different algorithms is not
constant for larger or smaller simulations.

We demonstrate the strong scalability of the \ao simulation platform
by keeping the simulation a fixed size, but increasing the number of
processors that are used.  Table.~\ref{tab:parallelsim} shows the
simulations parallelised by placing different simulation objects on
different processing nodes.  For the small simulation used for this
demonstration, this type of parallelisation can be sub-optimal,
because the processing load can be poorly balanced between processors.
For example, when placed on two processors, one of these will have two
wavefront sensor objects, requiring approximately double the
processing time of the other processor (with only one wavefront sensor
object).

\begin{figure}
\includegraphics[width=8cm]{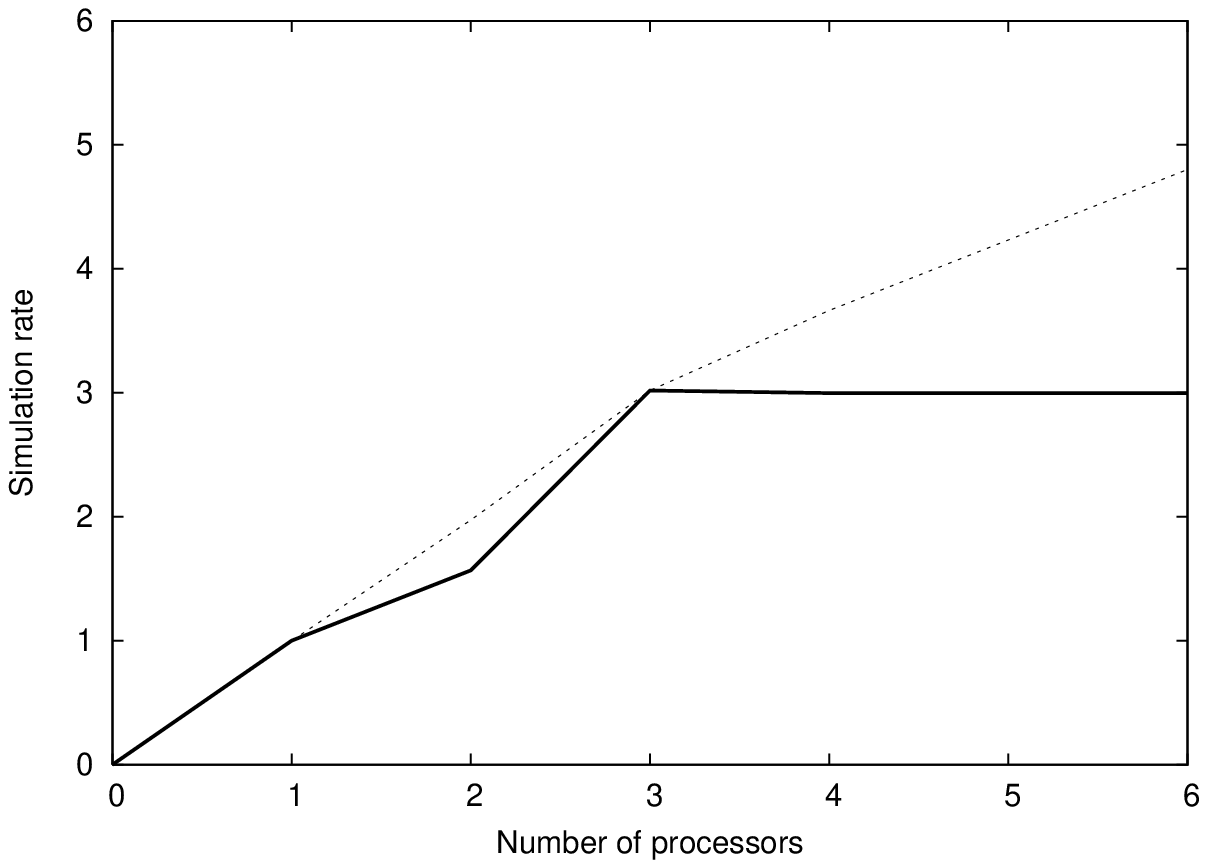}
\caption{A figure showing the number of simulation time-steps computed
  per unit time (simulation rate) when the simulation is parallelised
  over different numbers of processors.  The solid line shows the case
  when individual objects are not parallelised, while the dotted line
  shows the case when the Shack-Hartmann image creation and wavefront
  sensing algorithm is parallelised, providing a better fit to larger
  numbers of processors.  The simulation rate has been normalised to
  unity by the rate for an unparallelised simulation (with
  unparallelised simulation objects).}
\label{fig:timings}
\end{figure}

By parallelising some of the simulation objects (in this case the
wavefront sensing objects), the computational load can be spread more
evenly across processors, thus giving a better performance scaling
with computer system size.  Table.~\ref{tab:parallelsim2} shows the
simulations parallelised by using parallelised wavefront sensing
objects, allowing a better fit to a greater number of processors to be
realised as the processing load can be distributed more evenly.  The
timing results for these simulations are shown in
Fig.~\ref{fig:timings}.  This figure shows that the simulation can
scale well when it is well suited to the number of processors, for
example, using three processors gives a simulation rate three times
greater than one processor.  However, when the simulation is not well
suited to the number of processors (for example 2, 4, 5 and 6
processors in the case when the individual simulation objects are
unparallelised), the performance is sub-optimal.  If individual
objects are parallelised, the simulation can be fitted better to the
number of available processors, as the dotted line in
Fig.~\ref{fig:timings} shows.  However, currently, not all simulation
objects can be parallelised.

\begin{table}
\begin{tabular}{lll}\hline
Simulation object & Significant algorithms & Computation time / s\\ \hline
Infinite phase screen generation & Matrix operations & $10^{-4}$\\
Atmospheric pupil phase & Matrix operations & $7\times10^{-4}$\\
Deformable mirror simulation & Matrix operations & 0.03\\
Shack-Hartmann sensor, slope computation & FFT,
matrix operations & 0.18\\
Wavefront reconstruction (SOR) & Matrix operations & 0.02\\
Science image generation & FFT, matrix operations & 0.02\\ \hline
\end{tabular}
\caption{A table describing the simulation objects used in a study of
  the AO simulation strong scalability.}
\label{tab:scaleObjects}
\end{table}

\begin{table}
\scriptsize
\begin{tabular}{l|ll|lll|llll}\hline

ONE CPU CONFIGURATION & \multicolumn{2}{l|}{TWO CPUs} &
\multicolumn{3}{l|}{THREE CPUs} & \multicolumn{4}{l}{FOUR CPUs}\\ 
CPU1&CPU1&CPU2&CPU1&CPU2&CPU3&CPU1&CPU2&CPU3&CPU4\\ \hline
1. Phase screen (2km) & 1. & 4. & 2. & 4. & 1. & 2. & 1. & 4. & 5.\\
2. Phase screen (0km) & 2. & 5. & 3. & 7. & 5. & 3. & 3. & 7. & 8.\\
3. Telescope pupil phase (direction 1)& 3. & 7. & 6. & 10.& 8.&6.&6.&10.&11.\\
4. Telescope pupil phase (direction 2)&6.&8.&9.&12.&11.&13.&9.&&\\
5. Telescope pupil phase (direction 3)&9.&10.&13.&&&12.&&&\\
6. Deformable mirror (direction 1)&13.&11.&&&&&&&\\
7. Deformable mirror (direction 2)&12.&&&&&&&&\\
8. Deformable mirror (direction 3)&&&&&&&&&\\
9. Wavefront sensor (direction 1)&&&&&&&&&\\
10. Wavefront sensor (direction 2)&&&&&&&&&\\
11. Wavefront sensor (direction 3)&&&&&&&&&\\
12. Wavefront reconstructor&&&&&&&&&\\
13. Science calculation (science image)&&&&&&&&&\\ \hline
\end{tabular}
\normalsize
\caption{A table showing how simulation objects can be placed on
  different computing nodes to parallelise a simulation.  The first
  column gives a brief description of each object, the numbers of
  which are then referred to in other columns.  }

\label{tab:parallelsim}
\end{table}

\begin{table}
\scriptsize
\begin{tabular}{ll|ll|lll|llll|lll}\hline

\multicolumn{2}{l|}{ONE CPU} & \multicolumn{2}{l|}{TWO CPUs} &
\multicolumn{3}{l|}{THREE CPUs} & \multicolumn{4}{l|}{FOUR CPUs} &
\multicolumn{3}{l}{SIX CPUs}\\ 
\multicolumn{2}{c|}{CPU1}&CPU1&CPU2&CPU1&CPU2&CPU3&CPU1&CPU2&CPU3&CPU4&CPU1&CPU2&CPU3\\ \hline
1.&2.&1.&2.&2.&4.&1.&1.&2.&5.&10.c&1.&2.&5.\\
3.&4.&4.&5.&3.&7.&5.&3.&4.&8.&10.d&3.&4.&8.\\
5.&6.&7.&8.&6.&10.a&8.&6.&7.&11.a&11.d&6.&7.&11.a\\
7.&8.&3.&10.c&9.a&10.b&11.a&9.a&9.d&11.b&&9.a&10.a&11.b\\
9.a&9.b&6.&10.d&9.b&10.c&11.b&9.b&10.a&11.c&&9.b&10.b&\\
9.c&9.d&9.a&11.a&9.c&10.d&11.c&9.c&10.b&12.&&13.&&\\
10.a&10.b&9.b&11.b&9.d&12.&11.d&13.&&&&CPU4&CPU5&CPU6\\
10.c&10.d&9.c&11.c&13.&&&&&&&9.c&10.d&11.d\\
11.a&11.b&9.d&11.d&&&&&&&&9.d&10.d&11.d\\
11.c&11.d&10.a&12.&&&&&&&&&12.&\\
12.&13.&10.b&&&&&&&&&&&\\
&&13.&&&&&&&&&&&\\ \hline
\end{tabular}
\normalsize
\caption{A table showing how the parallelisation of simulation objects
  (denoted here by a, b, c and d suffixes for a four way
  parallelisation of the wavefront sensing algorithm) can be used to
  fit a simulation to a given number of CPUs.  The numbers represent
  the simulation objects described in table~\ref{tab:parallelsim}}
\label{tab:parallelsim2}
\end{table}

\subsection{ELT simulation suitability}
The Durham \ao simulation platform is suited for the simulation of ELT
scale \ao systems.  The XD1 supercomputer has 8~GB memory per
computing node, allowing large phase-screens, large numbers of
wavefront sensing elements (for example, Shack-Hartmann sub-apertures),
and other data to be stored.  The tight integration of the \fpgas with
memory and CPUs means parts of the simulation can be accelerated by
several orders of magnitude, and the high bandwidth, low latency
connections between nodes allows data to be passed rapidly between
parallelised processes.  This simulation platform provides the
capability for rapid simulation of \ao systems on all current
telescopes and next generation \elts.

\subsubsection{ELT simulation details}
A simulation of a classical \ao system on an \elt has been created to
demonstrate the use of the \ao simulation platform.  The key
parameters of this simulation are detailed in table~\ref{tab:eltsim}.
This simulation uses an infinite phase screen generator with von
Karman statistics \citep{assemat}.  A successive over-relaxation (SOR)
wavefront reconstructor is used, which means that it is not necessary
to create and invert an interaction matrix of the system. In a system
of this size, a full interaction matrix could easily take more memory
than available on our Cray XD1 system, also taking a prohibitive
length of time (days or weeks) to invert to obtain the control matrix,
and so conventional wavefront reconstruction is not an option.  We are
currently implementing sparse matrix algorithms and Fourier domain
wavefront reconstruction algorithms which will greatly reduce the
memory and computation requirements.  An \fpga hardware accelerator is
used for computation of the Shack-Hartmann images, and the spot
centroid location algorithm.  The high number of pixels per
sub-aperture allows elongated Shack-Hartmann spots (e.g.\ from a laser
guide star) to be analysed.  The simulation includes one wavefront
sensor, and one science target.  A more useful simulation may include
several wavefront sensors and several science targets, though these
are not presented here.

\begin{table}
\begin{tabular}{ll}\hline
Simulation parameter & Value\\ \hline
Telescope primary & 42m\\
Atmospheric layers & 3 (0~km, 2~km, 10~km)\\
Wavefront sensors & 1\\
Number of sub-apertures & $256\times256$ (16~cm per sub-aperture)\\
CCD pixels per sub-aperture & $16\times16$\\
Deformable mirrors & 1\\
Number of deformable mirror actuators & $256\times256$\\
Atmospheric resolution & 1~cm of sky per phase pixel\\
Phase pixels for science image creation & $4096\times4096$\\
Guide stars & 1 (natural guide star) \\ \hline
\end{tabular}
\caption{A table showing the ELT simulation model details}
\label{tab:eltsim}
\end{table}

This simulation has been parallelised over five nodes of the Cray XD1,
one node for each atmospheric layer, one node for the science target,
and one node to combine the atmospheric layers to give the atmospheric
phase at the telescope pupil, perform the simulation of the wavefront
sensor, and reconstruct the wavefront allowing the deformable mirror
surface to be reshaped.  Table~\ref{tab:elttimings} shows the relative
time spent computing each of these algorithms, and it can be seen that
by far the most computationally intensive is the simulation of the
science target (involving a $8192\times8192$ fast Fourier transform
for each simulation time-step).  These timings are pessimistic (worse
case), as they include computation of all scientific parameters,
including Strehl ratio and enclosed energy, which would typically only
be performed every hundred or so time-steps.  Without these
calculations, the science object takes less than 50 seconds to compute
and store the instantaneous point spread function for this \elt
simulation.  It should be noted that these timings do not scale in the
same way as a function of system size; for example when simulating a
smaller telescope, computation of the science image takes a
significantly smaller fraction of processor time.

For the majority of the time, the other processors are idle, waiting
for the science image algorithm to complete.  Work is currently
underway to place the bulk of this algorithm into hardware, which will
result in a significant performance improvement (a factor of 10 times
is expected).  The computation time of the science target simulation
currently scales as $O(n^2\log n)$, due to the large two dimensional
fast Fourier transform, where $n$ is the linear size of the phase
screen (measured in pixels).  This algorithm also uses the most memory
as it has to store a zero-padded pupil phase (so that the Fourier
transform is sampled at the Nyquist frequency), and both an
instantaneous and integrated point spread function.  The memory
requirements for this algorithm scale as $O(n^2)$ where $n$ is the
linear size of a phase screen, and over 5~GB memory were required for
this algorithm in the example here.  With the current hardware, it
would be possible to create a simulation with one more science object
(on the currently spare processing node), and about six more wavefront
sensing objects, to create (for example) a \mcao simulation, without
increasing the simulation iteration time.  This has not been
implemented at the present time, as the \mcao wavefront reconstructor
is not yet complete.

\begin{table}
\begin{tabular}{ll}\hline
Algorithm & Time taken / s\\ \hline
Science image and statistics & 70 \\
Atmospheric pupil phase  &6.1\\
Deformable mirror & 2.8\\
Wavefront sensing (Shack-Hartmann sensor, slope computation) & 0.6\\
Wavefront reconstruction (SOR) & 2.0\\
Phase screen generation & 2.9 per layer\\ \hline
\end{tabular}
\caption{A table showing the time spent in each algorithm of the ELT
  simulation.  The time taken for each simulation iteration
  (corresponding to 5~ms real time) was about 70 seconds, limited by
  the time to perform the science image computation.}
\label{tab:elttimings}
\end{table}

The planet finder instrument for the European Southern Observatory
\elt project is currently specified to have $200\times200$
sub-apertures \citep{myers}, and this is one of the most demanding
proposals.  The simulation demonstrated here is therefore of higher
order (has a larger number of sub-apertures) than all present and
planned astronomical \ao systems.

\section{Simulation results for ground layer adaptive optics}
A classical or single guide star \ao system can produce only a small
corrected field of view, and isoplanatic errors cause the image
quality to quickly degrade from the centre of this field.  When
natural guide stars are used, the sky coverage for these \ao systems
is severely limited, since it is difficult to find stars that are
bright enough within each isoplanatic patch of sky.  \Glao was
proposed as a solution to this problem, by applying a limited \ao
correction for a large field of view under any atmospheric conditions
at optical and infra-red wavelengths \citep{2002bcao.conf...11R}.  A
\glao system is not designed to produce diffraction limited images,
but improves the concentration of the \psf by correcting only the
lowest turbulent atmospheric layers.  Correction is then virtually
identical over the entire field of view since these layers are closer
to the ground, while the uncorrected higher layers degrade the spatial
resolution isoplanatically.

At Durham, we have implemented a \glao simulation model using the \ao
simulation framework for corrected fields up to 15 arc-minutes in size
based on high resolution turbulence profiles taken at the Gemini
observatory \citep{geminipasp}, and some of the results are presented
here to demonstrate an actual use of the simulation.  The Durham
simulation model includes detailed \wfs noise propagation and produces
2D \psfs, and is used to quantify the effects of such noise on the
\psf parameters across the \glao field for various seeing and noise
conditions.  The capabilities of this model are summarised:
\begin{enumerate}
\item The atmosphere can be modelled as any number of independently
  moving turbulent layers.
\item Multiple laser beacons and guide stars can be modelled.
\item Multiple deformable mirrors of different types can be modelled.
\item Multiple wavefront sensors can be included, encompassing all
  main detector noise effects, pixellation and atmospherically induced
  speckle.
\item The science \psf can be sampled at any number of field points
  simultaneously.  
\end{enumerate}

It is wholly-independent code (not derived from any other simulation
platform), but can be used subject to detailed cross-checks with other
\ao models \citep{geminipasp}.  This checking has been carried out as
part of work for the Gemini telescope consortium.  The simulation can
also be used for situations where the atmosphere cannot be treated as
stratified in layers, but as a three dimensional entity simply by
implementing such a model.  However, this is not considered here.

\subsection{Durham implementation}
\begin{figure}[htbp]
\centerline{\includegraphics[width=3cm]{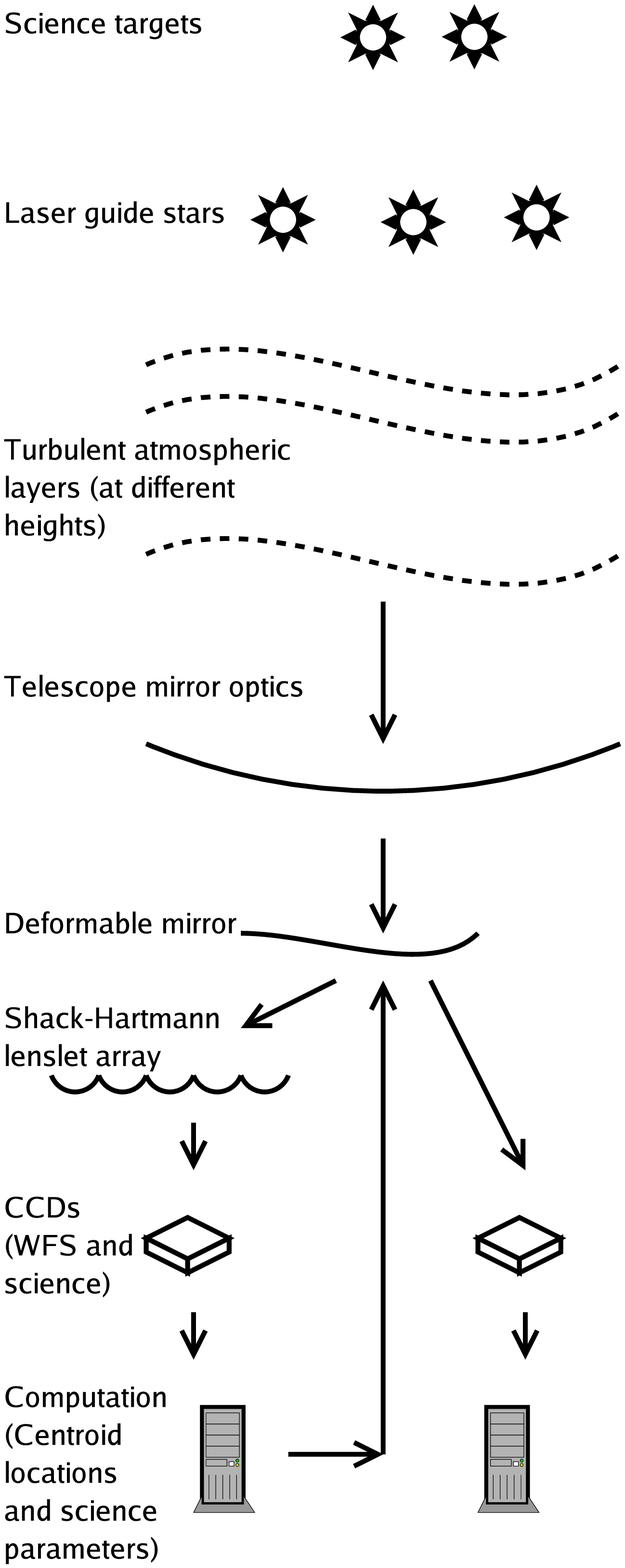}}
\caption{A figure showing the design of a GLAO system.}
\label{fig:glaodesign}
\end{figure}

A design for the \glao system is shown in Fig.~\ref{fig:glaodesign},
and this indicates that there are multiple guide stars and multiple
science sampling points where the \ao system performance is
categorised.

\begin{figure}[htbp]
\centerline{\includegraphics[width=6cm]{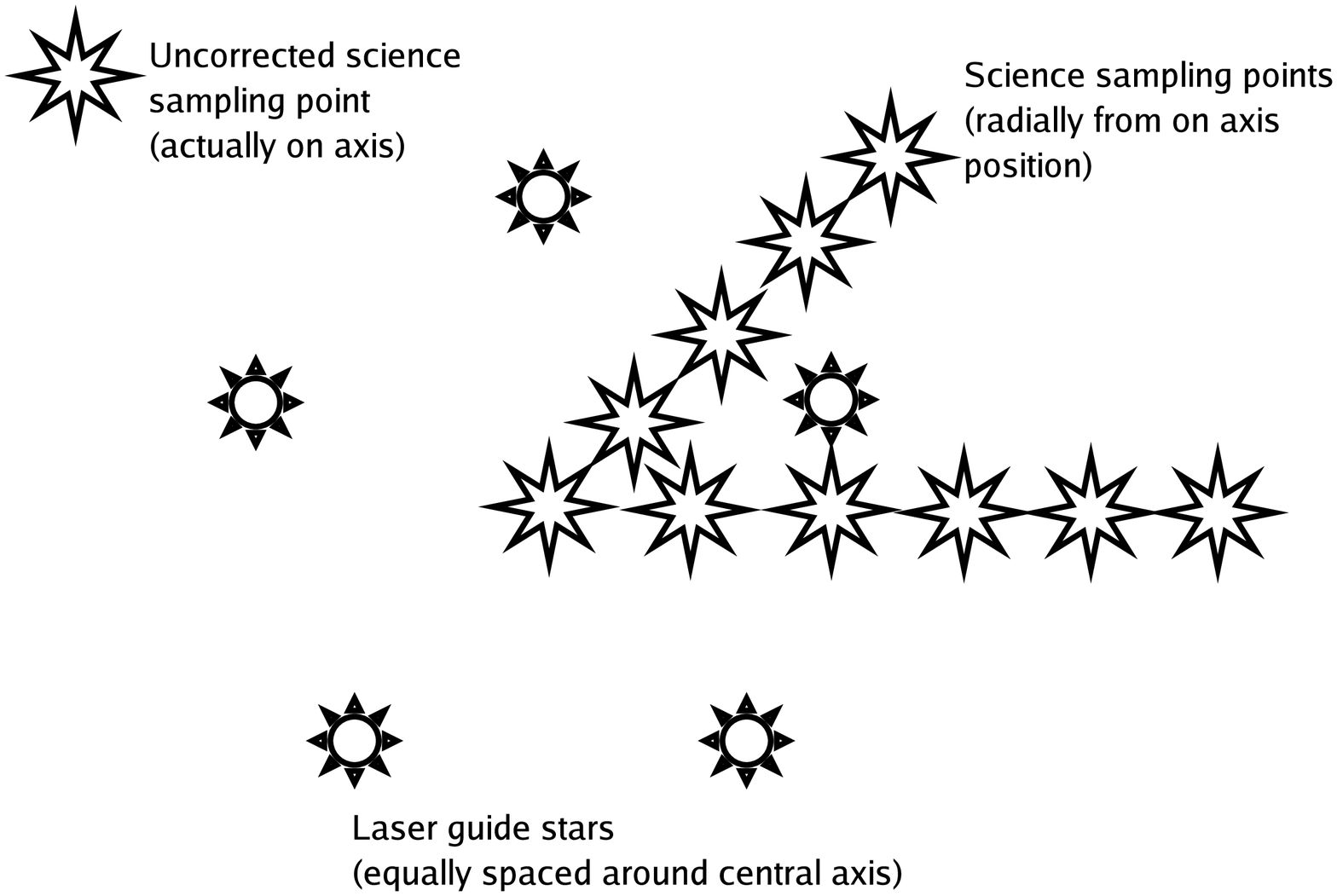}}
\caption{A schematic diagram of the relative positions of the laser
  guide stars and science sampling points used for the GLAO
  simulation.  The science sampling points (larger stars) are spaced
  uniformly 150~arc-seconds apart, while the laser guide stars
  (smaller stars) are positioned equally around a circle with a
  diameter which can be varied between 200--750~arc-seconds.}
\label{fig:glaolayout}
\end{figure}

We have simulated a system with five laser guide stars, and four
discrete atmospheric turbulence layers as shown in
table~\ref{tab:atmos}, assuming an 8~m telescope primary mirror.  The
simulation takes samples of the science field at a wavelength of
1250~nm at ten positions, on and off-axis, as well as the uncorrected
image, and uses these samples to categorise the performance of the \ao
system, with parameters such as the Strehl ratio and encircled energy
being computed for each science target location.  The simulation uses
a Shack-Hartmann wavefront sensor with $10\times 10$ sub-apertures,
and assumes a generic deformable mirror to which combinations of
Zernike modes are applied to give the correct mirror shape at each
time-step (the first 54 Zernike modes were corrected).  A typical
layout of the science stars and guide stars is shown in
Fig.~\ref{fig:glaolayout}, as viewed from the telescope.  The guide
star angle from the on-axis direction can be varied between
200--750~arc-seconds, and this can be used to investigate the degree
of \ao correction, and the area over which this correction is
achieved.  The integrated seeing in these models was taken as
0.6~arc-seconds with a Fried parameter of 0.17.  An exposure time of
100 seconds was used with a \wfs integration time of 2~ms.  The laser
guide stars were assumed to be of 13th magnitude brightness.
\begin{table}
\begin{tabular}{lllll}\hline
Layer height / m & 0 & 300 &  2000 & 10000 \\
Wind speed / ms$^{-1}$ & 6 & 9 & 10 & 18\\
Relative layer strengths & 0.45 & 0.15 & 0.07 & 0.33 \\ \hline
\end{tabular}
\caption{A table showing the atmospheric model details}
\label{tab:atmos}
\end{table}

\subsection{Parallelisation approaches}
\begin{figure}[htbp]
\centerline{\includegraphics[width=8cm]{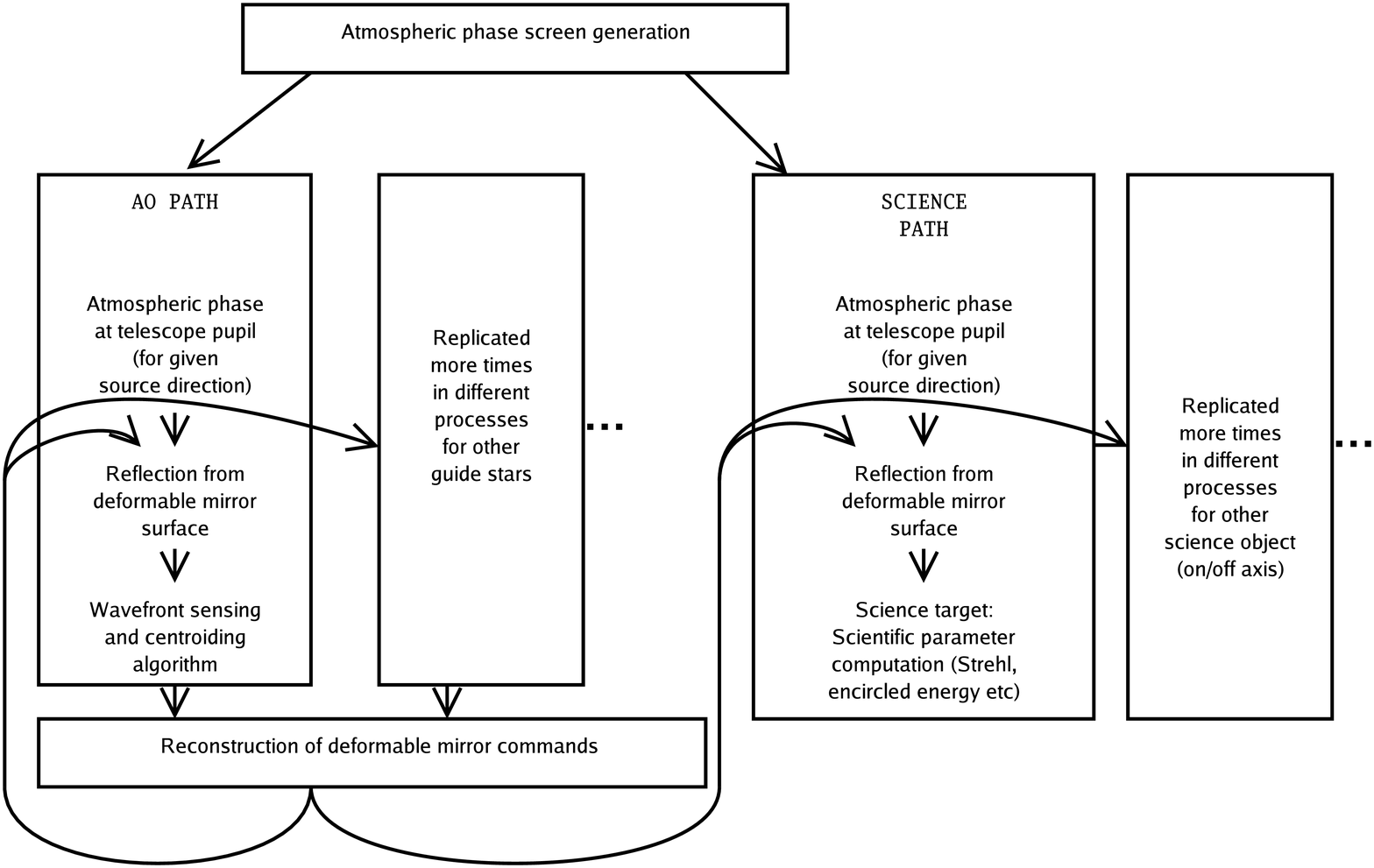}}
\caption{A flowchart showing how the GLAO simulation is carried out
  using the Durham AO simulation platform.  The algorithms in
  different boxes are implemented on different processors, and arrows
  show the direction of data flow between the algorithms.  Typically,
  there will be between 5 and 10 science paths (to determine how the AO
  performance changes at different angles from the vertical axis), and
  between 5 and 10 AO paths, depending on the number of laser guide stars
  being used.}
\label{fig:glaosim}
\end{figure}

There are many ways in which a large simulation such as that presented
here can be parallelised.  The optimal parallelisation approach will
reduce the bottlenecks in data transferred between processes and
minimise the amount of time in which processors are not actively
processing, whilst fully utilising as many processors as possible.  As
mentioned in section~\ref{sect:parallel}, when simulating an \ao
system, it is possible to separate the parallel optical paths from
different guide stars and science targets on to different processors,
reducing the data transfer between processes.  This is the approach
used here, and is presented as a flowchart in Fig.~\ref{fig:glaosim}.

\subsection{GLAO simulation results}
When a number of guide stars are evenly spaced about a circle (as
viewed from the telescope), there will be some atmospheric correction
for starlight passing within this circle, but the degree of correction
will fall for starlight outside the circle.  If the guide star
separation is reduced, better correction will be achieved over a
smaller area.  Conversely, if the separation is increased, a poorer
correction will be achieved over a larger area.

A \glao system does not aim to achieve a high degree of correction.
Rather, a partial degree of correction is achieved over a wide field
of view, and the \glao system is usually designed to be complementary
to more conventional \ao systems, or to be used with integral field
spectroscopy units.  The correction achieved from a \glao system alone
typically produces Strehl ratios of only a few percent.

\begin{figure}[htbp]
\centerline{\includegraphics[width=8cm]{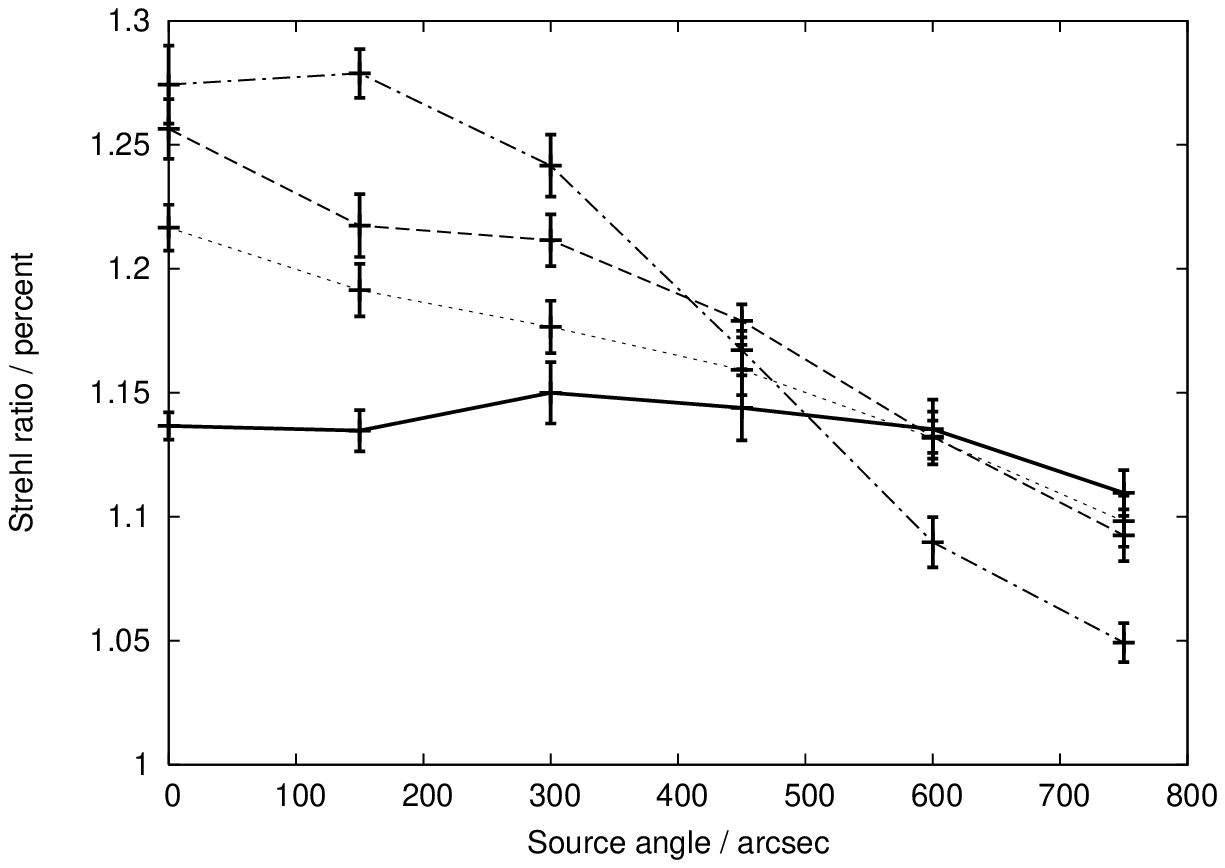}}
\caption{A figure showing the Strehl ratio as a function of distance
  from the on-axis direction.  The solid line is for a laser guide
  star separation of 700 arc-seconds from the on-axis direction, the
  dotted line for a separation of 550 arc-seconds, the dashed line for
  a separation of 450 arc-seconds and the dot-dashed line for a
  separation of 250 arc-seconds.  The points and error bars are
  obtained from a sample of 10 results for each point.}
\label{fig:glaostrehl}
\end{figure}

The results of an investigation into the effect of guide star
separation are presented here, and Fig.~\ref{fig:glaostrehl} shows
that by moving the guide stars out from the on-axis direction, the
isoplanatic correction covers a larger area with a smaller degree of
correction, hence giving a lower Strehl ratio.  This decreases for
fields further from the on-axis direction, but the rate of change is
dependent on the guide star separation.  The uncorrected Strehl ratio
was about 0.75~percent.

\begin{figure}[htbp]
\centerline{\includegraphics[width=8cm]{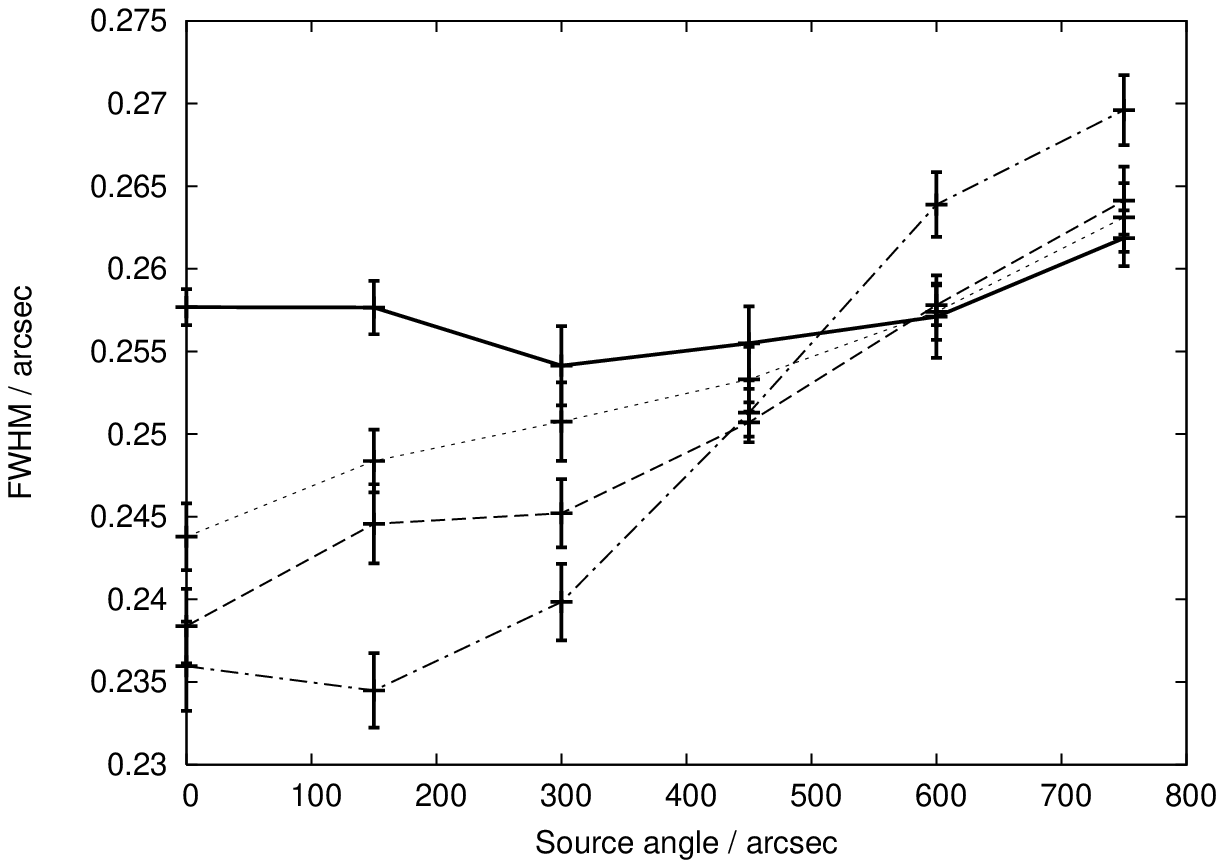}}
\caption{A figure showing the change in the FWHM of corrected images
  as a function of angle from the on-axis direction.  The solid curve is
  calculated with a guide star angle of 700 arc-seconds from the axis,
  the dotted curve for 550 arc-seconds, the dashed curve for 450
  arc-seconds and the dot-dashed curve for 250 arc-seconds.}
\label{fig:glaoFwhm}
\end{figure}

The \fwhm as a function of angle from the on-axis direction also
displays the expected behaviour, increasing as the viewing angle is
moved away from the axis.  When the guide star separation is small,
the \fwhm is small close to the axis, increasing rapidly away from it,
and when guide star separation is large, the \fwhm is initially
larger, but increases slowly away from the on-axis direction, as shown
in Fig.~\ref{fig:glaoFwhm}.  The uncorrected \fwhm was 0.35~arcsec.

The \glao simulation presented here has been compared with other
independent \ao simulation codes \citep{geminipasp}, and are found to
be in agreement within the statistical uncertainties.

\section{Conclusion}
We have developed a new \ao simulation capability at Durham for
astronomical applications, and this platform is capable of extremely
large telescope \ao system simulation.  This simulation platform is
capable of using algorithms implemented within reconfigurable logic to
provide hardware acceleration for the most computationally intensive
tasks.  

A simulation platform includes tools for creating and controlling the
simulations, and optimal parallelisation techniques specific to \ao
simulations have been discussed.  The flexibility of the simulation
platform, as well as the ability to query and alter the state of a
running simulation make it unique.  Additionally, techniques used to
parallelise a given simulation reducing the computation time have been
described, and these parallelisation strategies are specifically aimed
at \ao system simulation.  The simulation platform has been tested
against other independent codes, and is found to be in agreement with
these.

We have demonstrated a use of the \ao simulation platform for \glao
simulation, and presented some results obtained.  These results show
that separation of guide stars affects the achievable \ao correction
and the area over which this correction can be achieved.


\end{document}